\documentclass[twocolumn,superscriptaddress]{revtex4}
\usepackage{amsmath}
\usepackage{amssymb}
\usepackage{graphicx}
\usepackage[load-configurations = abbreviations]{siunitx}
\usepackage{xcolor}

\begin{document}

\title{Lattice-Shifted Nematic Quantum Critical Point in FeSe$_{1-x}$S$_x$}
\author{S. Chibani}
\author{D. Farina}
\author{P. Massat}
\author{M. Cazayous}
\author{A. Sacuto}
\affiliation{Universit\'e de Paris, Mat\'eriaux et Ph\'enom\`enes Quantiques, CNRS, F-75205 Paris, France}
\author{T. Urata}\thanks{Present address : Department of Materials Physics, Nagoya University, Chikusa-ku, Nagoya 464-8603, Japan}
\affiliation{Department of Physics, Graduate School of Science, Tohoku University, Sendai 980-8578, Japan}
\author{Y. Tanabe}
\affiliation{Department of Physics, Graduate School of Science, Tohoku University, Sendai 980-8578, Japan}
\affiliation{Department of Applied Science, Okayama University of Science, Okayama, 700-0005, Japan}
\author{K. Tanigaki}
\affiliation{Department of Physics, Graduate School of Science, Tohoku University, Sendai 980-8578, Japan}
\affiliation{Advanced Institute for Materials Research, Tohoku University, Sendai 980-8577, Japan}
\author{A. E. B\"ohmer}
\affiliation{Karlsruhe Institute of Technology, Institute for Quantum Materials and Technologies, 76021 Karlsruhe, Germany}
\affiliation{Ames Laboratory US DOE, Iowa State University, Ames, Iowa 50011, USA}
\author{P. C. Canfield}
\affiliation{Ames Laboratory US DOE, Iowa State University, Ames, Iowa 50011, USA}
\affiliation{Department of Physics and Astronomy, Iowa State University, Ames, Iowa 50011, USA}
\author{M. Merz}
\affiliation{Karlsruhe Institute of Technology, Institute for Quantum Materials and Technologies, 76021 Karlsruhe, Germany}
\author{S. Karlsson}
\author{P. Strobel}
\author{P. Toulemonde}
\affiliation{Universit\'e Grenoble Alpes, CNRS, Grenoble INP, Institut N\'eel, F-38000 Grenoble, France}
\author{I. Paul}
\affiliation{Universit\'e de Paris, Mat\'eriaux et Ph\'enom\`enes Quantiques, CNRS, F-75205 Paris, France}
\author{Y. Gallais}\email{yann.gallais@u-paris.fr}
\affiliation{Universit\'e de Paris, Mat\'eriaux et Ph\'enom\`enes Quantiques, CNRS, F-75205 Paris, France}

\date{}


\begin{abstract}
We report the evolution of nematic fluctuations in  FeSe$_{1-x}$S$_x$ single crystals as a function of Sulfur content $x$ across the nematic quantum critical point (QCP) $x_c\sim$ 0.17 via Raman scattering. The Raman spectra in the $B_{1g}$ nematic channel consist of two components, but only the low energy one displays clear fingerprints of critical behavior and is attributed to itinerant carriers. Curie-Weiss analysis of the associated nematic susceptibility indicates a substantial effect of nemato-elastic coupling which shifts the location of the nematic QCP. We argue that this lattice-induced shift likely explains the absence of any enhancement of the superconducting transition temperature at the QCP. The presence of two components in the nematic fluctuations spectrum is attributed to the dual aspect of electronic degrees of freedom in Hund's metals, with both itinerant carriers and local moments contributing to the nematic susceptibility.
\end{abstract}
\maketitle
\section*{Introduction}

The link between quantum criticality and the emergence of unconventional superconducting (SC) states is ubiquitous among several families of materials including heavy-fermion, cuprates and iron-based (Fe SC) superconductors \cite{Lohneysen2007,Shibauchi2014,Taillefer2010}. The presence of divergent critical fluctuations associated to a nearby order are thought to provide a pairing glue and significantly enhance the superconducting transition temperature \cite{Monthoux2007,Metliski2015}. Dating from the seminal work of Berk and Schrieffer \cite{Berk1966}, most studies have been devoted to fluctuations near a magnetic instability, and in particular antiferromagnetic (AF) ones which give rise to non s-wave pairing states such as d-wave in the cuprates, and possibly in heavy fermion and organic SC \cite{Moriya-Ueda2000}.

In many Fe SC, SC emerges on the border of an AF state that is however also accompanied, or even preceded, by an electron nematic phase whereby the electron fluid spontaneously breaks the four-fold rotational symmetry of the underlying tetragonal lattice \cite{Fradkin2010,Fernandes-review}. The initial focus was on the proximity of the AF phase which is expected to lead a sign changing s$^{\pm}$ SC pairing state \cite{Kuroki2008,Mazin2008}, but recently the role of nematic fluctuations and criticality on the SC state has come under scrutiny \cite{Chu2012,Fernandes2013,Gallais2016a,Hosoi2016,Kuo2016,Wang2018,Klein2019,Eckberg2020,Hong2020}. Several theoretical works have argued that the SC pairing is generically enhanced near a nematic quantum critical point (QCP)\cite{Yamase2013,Maier2014,Lederer2015,Metliski2015,Labat2017}, even if it is not the leading pairing glue. On the experimental side, disentangling the role of AF and nematic QCP has proved challenging because the associated orders are essentially concomitant in most Fe SC.  An exception is FeSe, where at ambient pressure SC emerges at about 8K out of a non-magnetic nematic state that sets in at much higher temperature $\sim$ 90K \cite{Boehmer-review, Coldea-review}. The origin of the nematic state in FeSe is currently under debate, with both orbital and spin degrees of freedom possibly playing a role \cite{Glasbrenner2015,Wang2015,Chubukov2016,Kontani2016,Fanfarillo2018}. Weakening the nematic order in order to reach a putative QCP can in principle be achieved by either physical or chemical pressure. In the case of hydrostatic pressure a magnetic order sets in before the end point of the nematic phase precluding a nematic QCP \cite{Bendele2010,Sun2016,Kothapalli2017,Massat2018}. By substituting selenium (Se) with isovalent sulfur (S) however, the nematic order can be continuously suppressed without any magnetic order \cite{Watson2015,Urata2016,Hosoi2016} (see fig. 1(a)), providing a model system to study the impact of a nematic QCP on SC and normal state properties.

Several studies indicate a significant impact of the nematic order on the SC pairing state \cite{Sprau2017,Hanaguri2018,Sato2018,Benfatto2018}. However, intriguingly the evolution of $T_c$ in FeSe$_{1-x}$S$_x$ indicates a marginal role for nematic quantum critical fluctuations in boosting SC. In fact coming from the tetragonal, non nematic side of the FeSe$_{1-x}$S$_x$ phase diagram, $T_c$ is essentially flat upon approaching the nematic QCP at $x_c\sim$0.17 showing only a mild maximum well-inside the nematic ordered phase \cite{Reiss2017,Wiecki2018} (Fig. 1(a)). This apparent contradiction with theoretical expectations was recently argued to arise from the coupling between electronic nematic degrees of freedom and the lattice which cuts-off nematic quantum criticality at low temperature. This effect may suppress the expected enhancement of $T_c$ found in electronic-only models \cite{Labat2017}. The lattice may also play a significant role in restoring Fermi liquid behavior of the normal state near the QCP \cite{Paul2017,Carvalho2019,Reiss2020}. In general the effect of the lattice appears to have been overlooked in several previous studies of the nematic QCP FeSe$_{1-x}$S$_x$ \cite{Hosoi2016,Licciardello2019}, calling for an experimental clarification of its potential role in the properties of FeSe$_{1-x}$S$_x$ near the nematic end-point.

\begin{figure*}
\includegraphics[width=2\columnwidth]{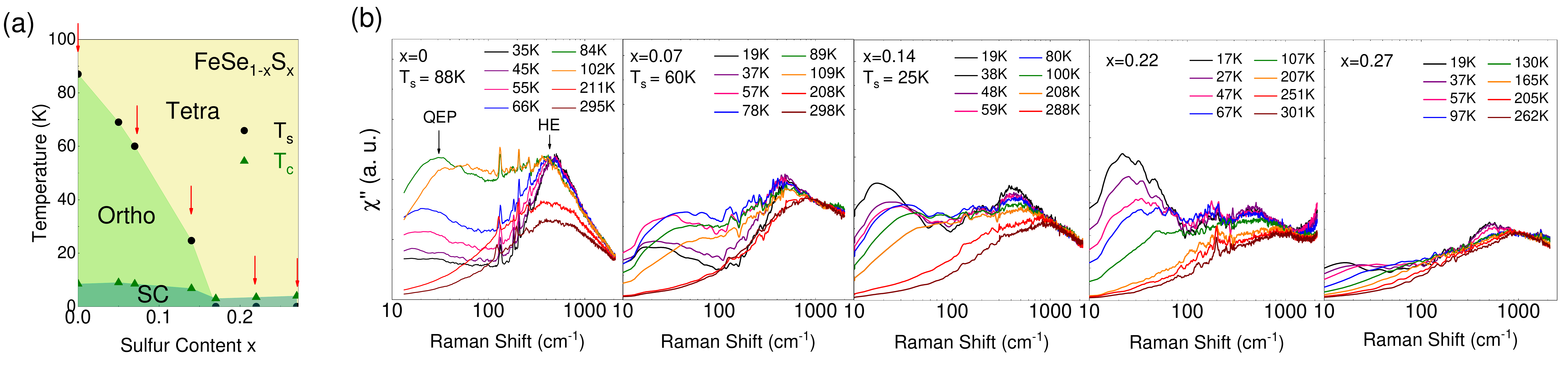}
\caption{\label{fig1} {\bf Raman spectra in the $B_{1g}$ nematic channel of FeSe$_{1-x}$S$_x$} (a) Phase diagram of FeSe$_{1-x}$S$_x$. The arrows indicates the compositions whose Raman spectra are shown in (b). (b) Raman response as a function of temperature in $B_{1g}$ symmetry channel for five different sulfur compositions. The energy is shown on a logarithmic scale. The quasi-elastic peak (QEP) and high-energy peak (HE) are indicated by arrows in the $x$=0 spectra.}
\end{figure*}
\begin{figure}
\includegraphics[width=1\columnwidth]{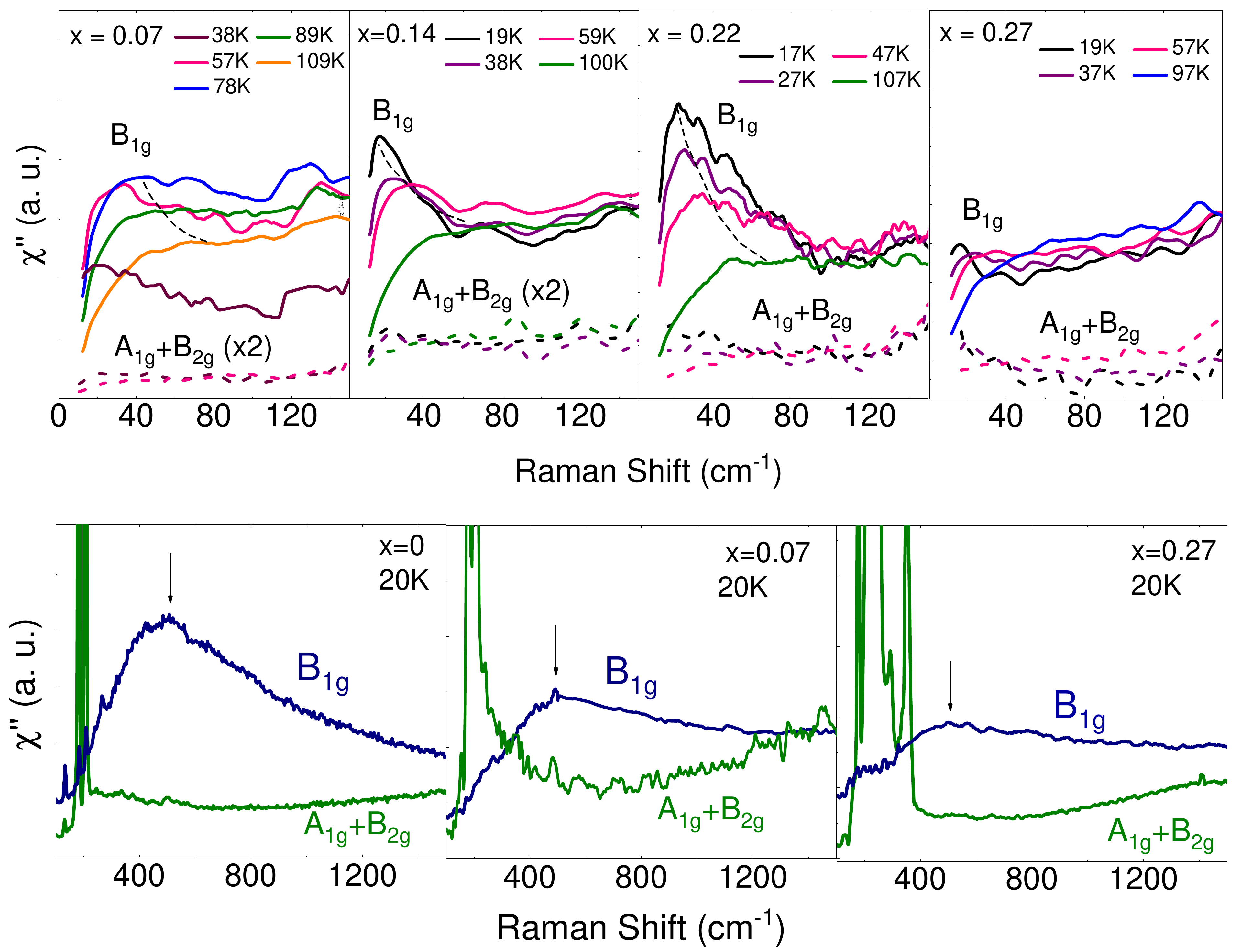}
\caption{\label{fig1b} {\bf Symmetry dependence of the Raman spectra}  Upper panel: Low energy part (on a linear scale) of the Raman spectra for 2 different symmetry channels $B_{1g}$ (full lines) and $A_{1g}$+$B_{2g}$ (dotted lines) and for $x$=0.07, $x$=0.14, $x$=0.22 and $x$=0.27. Lower panel: symmetry channel dependence of the spectra over a wide energy range on a linear scale for $x$=0, $x$=0.07 and $x$=0.27.}
\end{figure}


Here we use the ability of Raman scattering to probe symmetry resolved electronic fluctuations in the long wavelength limit ($q$=0) \cite{Yamase2011,Gallais2013,Gallais2016b,Thorsmolle2016,Massat2016,Auvray2019,Adachi2020} to investigate the evolution of nematic fluctuations in FeSe$_{1-x}$S$_x$  as a function of doping and temperature. Our study spans a significant portion of the phase diagram of FeSe$_{1-x}$S$_x$ , from $x$=0 to $x$=0.27. This allows us to assess the evolution of critical nematic fluctuations across the nematic QCP, located at $x_c\sim$0.17 \cite{Reiss2017,Hanaguri2018}. The dynamical nematic fluctuations spectrum consists of two main components, but only one of them shows a clear critical behavior upon approaching the nematic phase transition. A Curie-Weiss analysis of the critical component reveals a significant shift between the bare electron-only nematic QCP that is captured by Raman scattering, and the thermodynamic one. We attribute this shift to electron-lattice coupling, whose energy scale is found to be a sizable fraction of the typical Fermi energy of S-FeSe, and may explain the absence of $T_c$ enhancement at the nematic QCP. Our study highlights the important role of electron-lattice coupling effects in both superconducting and normal state properties of FeSe, and more generally of Fe SC.

\section*{Results}

\subsection*{Sulfur doping dependence of symmetry-resolved Raman spectrum}
In our study, seven different sulfur (S) compositions $x$ were studied. The S content of the studied single crystals, which usually differs significantly from the nominal one, was determined by element specific Energy Dispersive X-ray Spectroscopy (EDS) and single-crystal X-Ray Diffraction (XRD) measurements (see Methods). The structural /nematic transition $T_s$ was determined by transport measurements on crystals from the same batch, and also in-situ via the observation of the onset of elastic light scattering by nematic domain formation below $T_s$ (see supplementary fig. 1 \cite{SM}). $T_c$ was determined by the onset of diamagnetic signal in SQUID magnetometry. The phase diagram in Fig. 1 summarizes the properties ($x$, $T_s$, $T_c$) of the crystals studied. Details on the Raman scattering measurements can be found in the Method section.
\par
In Fig. 1b we show the temperature dependence of the Electronic Raman Scattering (ERS) spectrum for 5 different dopings (x=0, $x$=0.07(2), $x$=0.14(2), $x$=0.22(2) and $x$=0.27(2)) over a relatively wide energy range (0 - 2000 cm$^{-1}$, corresponding to 0-250 meV). Data for $x$=0 (FeSe) were already reported in \cite{Massat2016}. Incoming and outgoing photon polarizations were oriented so as to probe the $B_{1g}$ symmetry channel (using the 1 Fe unit cell notation), which transforms as $x^2$-$y^2$ and thus corresponds to the nematic order parameter symmetry (see Methods). For all compositions, two main features, marked by arrows in the $x$=0 spectra, can be distinguished: a relatively broad peak at high-energy (hereafter labeled HE peak) located at around 400 cm$^{-1}$ (50 meV), and a narrower feature, a quasi-elastic peak (QEP), located at much lower energy, below 100cm$^{-1}$ ($\sim$ 12 meV) at low temperatures. The low $x$ content data are consistent with a previous Raman study which reported data in a more limited spectral and doping range \cite{Zhang2017}. The HE peak intensity decreases with doping (see also fig. 2) and is only mildly affected by $T_s$. The QEP by contrast, is strongly affected by $T_s$ and displays a significant doping and temperature dependence : its intensity is maximum close to $T_s$ for lower $x$ compositions, and continuously increases down to the lowest measured temperature (17K) for $x$=0.22. The enhancement and collapse of the QEP across the nematic/structural transition temperature is consistent with previous studies in various Fe SC \cite{Gallais2013,Thorsmolle2016,Massat2016,Adachi2020}, and is ascribed to critical nematic fluctuations near $T_s$. The QEP intensity is significantly reduced for $x$=0.27 (see fig. 1b and 2) indicating a reduction of nematic fluctuations for high $x$ content. Figure 2 shows the symmetry dependence of the QEP and HE peak: consistent with the nematic fluctuations interpretation, the QEP is absent in both $B_{2g}$ and $A_{1g}$ channels throughout the compositional range studied. Interestingly, the HE peak display a robust symmetry dependence: it is observed only in the $B_{1g}$ channel for all $x$.

\begin{figure}
\includegraphics[width=1.0\columnwidth]{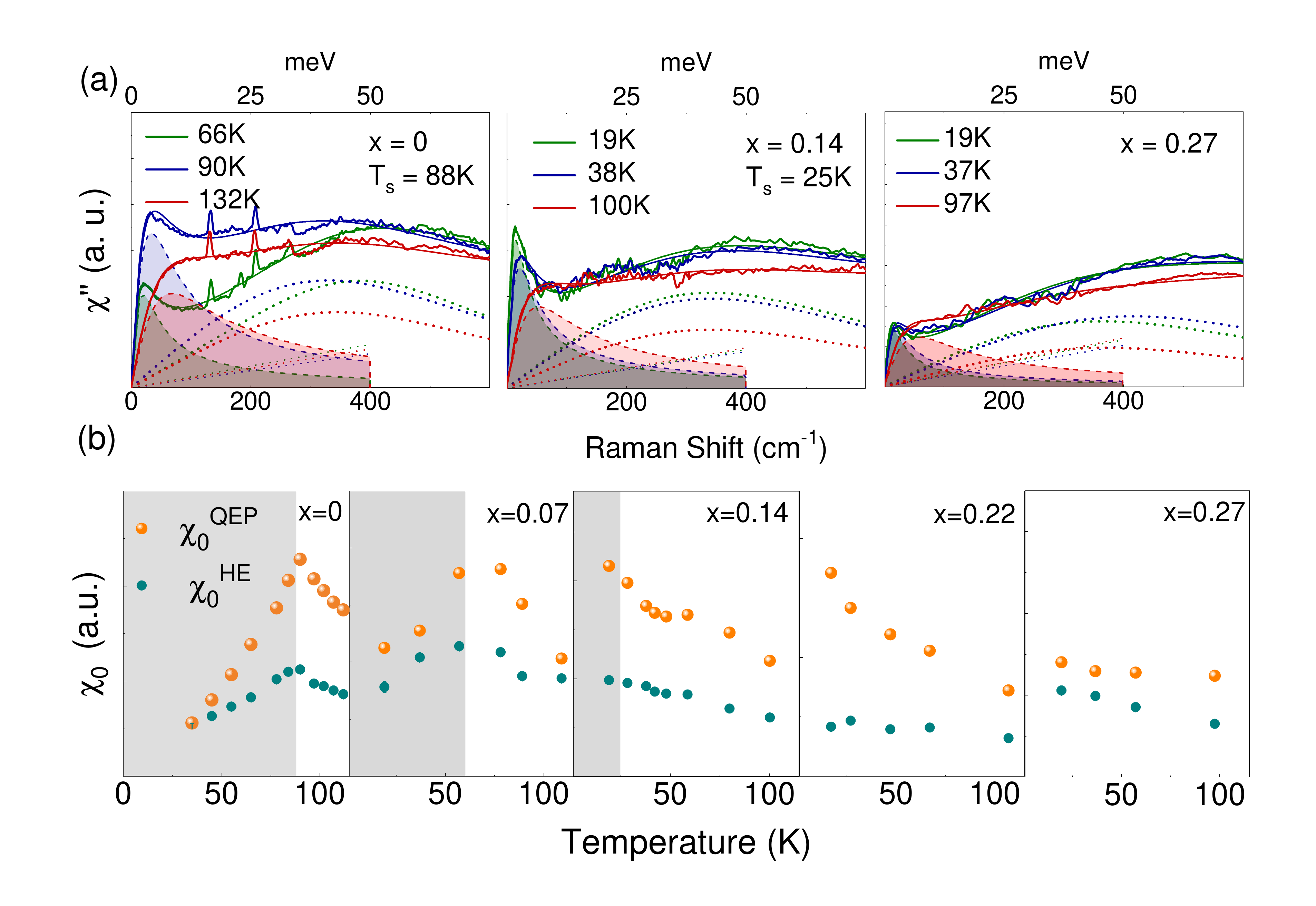}
\caption{\label{fig2} {\bf Two component analysis of the nematic fluctuation spectrum} (a) Decomposition of the $B_{1g}$ Raman response illustrated at 3 different compositions: $x$=0, $x$=0.14 and $x$=0.27. (see equation in the main text). The QEP peak component is highlighted via shading. (b) Temperature dependence of the two main contributions (QEP and HE) to the static nematic susceptibility $\chi_0$. Standard errors of the fits are shown but are usually smaller than the symbols.}
\end{figure}

\subsection*{Extracting critical nematic fluctuations}
In order to analyze in more details the two main components of the nematic fluctuation spectrum, we performed a two component fit of the Raman spectra, where the QEP and HE peaks are fitted by an overdamped Lorentzian and a damped oscillator respectively. $\chi''(\omega)= A\frac{\omega\Gamma_{QEP}}{\omega^2+\Gamma_{QEP}^2}+B\frac{\omega\omega_{HE}^2\Gamma_{HE}}{(\omega\Gamma_{HE})^2+(\omega^2-\omega_{HE}^2)^2}$. A small linear background was added to model the high energy part of the spectra. We discuss the possible origin the HE peak below. The overdamped functional form taken for the QEP can be linked to a Drude-like response of itinerant carriers that is renormalized near a nematic instability. In this framework $\Gamma_{QEP}$ is an effective quasiparticle scattering rate where the impurity scattering rate $\Gamma_0$ in the $B_{1g}$ channel is renormalized by the nematic correlation length $\xi_n$: $\Gamma_{QEP}$=$\Gamma_0(\frac{a}{\xi_n})^2$ where $a$ is the lattice parameter of the Fe plane \cite{Gallais2016b,Udina2020}.
 As shown in fig. 3a (see supplementary fig. 3 \cite{SM} for more doping and temperatures), all the spectra below $\sim$ 150K could be reproduced with this decomposition. At higher temperatures the QEP and HE essentially merge, making it difficult to unambiguously separate them via the fitting procedure. The HE peak energy was found to be weakly dependent of doping and temperature, but its width $\Gamma_{HE}$ displays a marked decrease below $T_s$ (see supplementary figure fig. 5 \cite{SM}). From the decomposition we can extract the behavior of the static nematic susceptibility $\chi_0$ associated to each component (fig. 3b). Indeed using Kramer-Kronig relations, we have: $\chi_0^{QEP}=\int_0^{\infty}\frac{\chi''}{\omega}d\omega$=A and equivalently $\chi_0^{HE}$=B. For all dopings the QEP component is the dominant contribution to the nematic susceptibility. Besides, as shown in fig. 3b only the QEP component shows clear fingerprint of critical behavior as function of temperature. An exception is $x$=0.27 where both contributions display a similar mild temperature dependence. This points to two distinct sources of nematic fluctuations in FeSe, with only one displaying fingerprints of critical behavior near the nematic QCP.

\begin{figure}
\includegraphics[width=1.0\columnwidth]{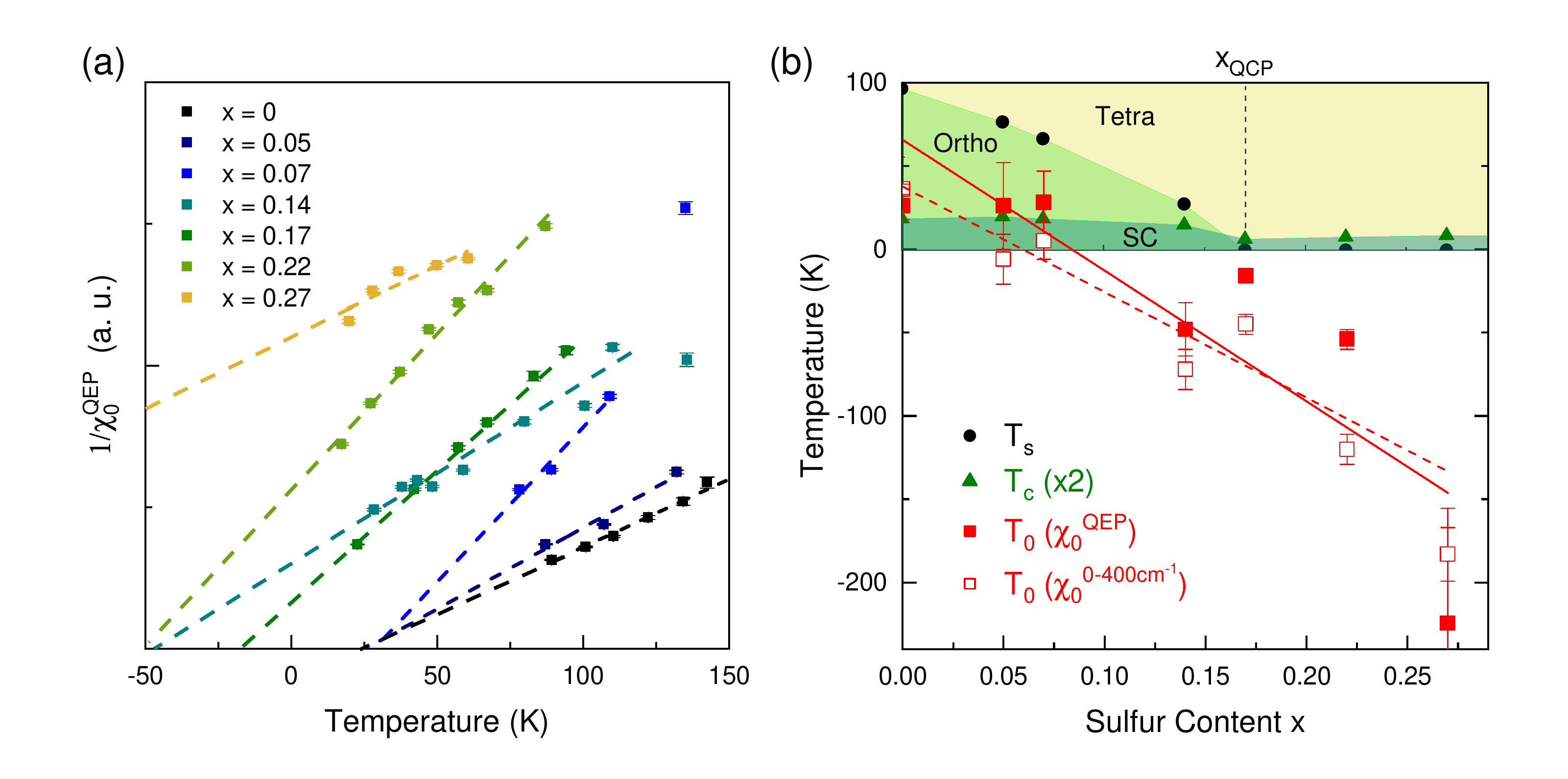}
\caption{\label{fig3} {\bf Curie-Weiss analysis of the critical nematic susceptibility} (a) Curie-Weiss analysis of the critical inverse nematic susceptibility component $\chi_0^{QEP}$. Note that only data above $T_s$ are shown for low $x$ compositions. (b) Phase diagram $x-T$ of FeSe$_{1-x}$S$_x$ showing the evolution of the Curie-Weiss nematic transition temperature $T_0$ extracted from the analysis in (a) (full square) The error bars for $T_0$ correspond to standard errors of the Curie-Weiss fits. Also shown in open squares are the $T_0$ values extracted from a Curie-Weiss analysis of the $\chi_0$ extracted without any fitting by performing a Kramers-Kronig integration of the raw data restricted to energies below 400 cm$^{-1}$ (see supplementary fig. 6 \cite{SM}). The lines are linear fits of $T_0(x)$ values.}
\end{figure}

\par
 We note that a different decomposition of the Raman response of FeSe was proposed in \cite{Zhang2017}, where a gap-like suppression was added in order to reproduce the loss of intensity of the spectra at low energy in $B_{1g}$ symmetry below $T_s$. In our case we found that the spectra across $T_s$ could be well reproduced without invoking such a gap. Rather the apparent spectral gap below $T_s$ results in our case from a combination of the collapse of the QEP contribution in the nematic phase and a sizable reduction of the width of the HE peak. A suppression of the nematic fluctuations, and thus the QEP, is qualitatively expected in the nematic ordered phase. Still the strong suppression of the QEP observed could also be reinforced by an orbital transmutation effect below $T_s$, whereby the hole pocket becomes essentially single component in orbital space inducing strong vertex corrections that further suppress the $B_{1g}$ Raman response in the nematic phase \cite{Udina2020}.
 \par
 Focusing now on the critical behavior of the QEP component, we display in fig. 4a $\frac{1}{\chi_0^{QEP}}$ as a function of $x$ and $T$. We have added two additional compositions where spectra were taken in a more limited spectral range, allowing only the extraction of the QEP component (see supplementary fig. 4 \cite{SM}). Assuming a mean-field Curie-Weiss temperature dependence, we can extract the corresponding nematic Curie-Weiss temperature $T_0$ whose $x$ dependence is shown in fig. 4b. The evolution of $T_0$ extracted directly from the raw data using Kramer-Kronig relations restricted to energies below 400 cm$^{-1}$ (labeled $T_0^{400cm^{-1}}$) is also shown for comparison (see supplementary fig. 6 \cite{SM}). While less transparent, this determination has the advantage of not relying on any assumption about the functional form of the low energy part of the spectrum. Consistently with low energy QEP part of the $B_{1g}$ spectrum being more critical, $T_0^{400cm^{-1}}$ tend to be lower than $T_0^{QEP}$ determined from the QEP part only. Note also that even lower $T_0$ are obtained if the full temperature dependent part of the spectra, thus including both QEP and HE contributions to $\chi_0$, is integrated from 0 to 2000 cm$^{-1}$ (see supplementary figure 6 and 7 \cite{SM}). Since our focus is on the impact of nematic criticality on low energy physics which governs both SC and transport, we believe that it is the behavior of the low energy critical QEP that is the most relevant for the present discussion. Therefore we restrict our discussion to $T_0^{QEP}$ and $T_0^{400cm^{-1}}$ below.
 \par
For both determinations, $T_0$ decreases with $x$ tracking $T_s$ but staying significantly below it (Fig. 4b).  The difference between $T_s$ and $T_0$ is a consequence of the finite electron-lattice coupling as first discussed by Kontani et al. \cite{Kontani2014}. Because the nematic susceptibility $\chi_0$ extracted from Raman spectra is in the dynamical limit (k=0 and then $\omega$=0, while the opposite is required for a thermodynamical susceptibility), the symmetry allowed coupling between electronic nematic fluctuations and the soft orthorhombic acoustical phonon vanishes, leaving only the contribution from the bare electronic-only nematic susceptibility \cite{Gallais2016b}. In this picture $T_0$ represents the mean-field nematic transition temperature in the absence of the lattice. By contrast in the opposite thermodynamical or static limit, the electron-lattice comes into play and stabilizes the nematic transition at $T_s > T_0$. We note that such a picture was validated by comparing the temperature dependence of the Raman nematic susceptibility and the shear modulus in both BaFe$_{1-x}$Co$_x$)$_2$As$_2$ and FeSe$_{1-x}$S$_x$ \cite{Gallais2016b,Massat2016}. In FeSe$_{1-x}$S$_x$ the difference between $T_s$ and $T_0$ is significant for the 4 compositions showing nematic order, ranging from 45K to 75K, depending on the method used to extract $T_0$. An unavoidable consequence of this large difference is a significant difference in the location of the associated critical points: the bare electronic nematic QCP extrapolates at $x^0_{QCP} \sim 0.07(\pm 0.04)$, but the thermodynamical QCP is located at $x_{QCP} \sim 0.17(\pm 0.02)$.  We stress that the bare nematic QCP is only fictitious. Still, its location is paramount since as we discuss below, it ultimately governs the impact of the critical nematic fluctuations on low energy properties such as SC pairing and transport lifetime.

\begin{figure}
\includegraphics[width=1.0\columnwidth]{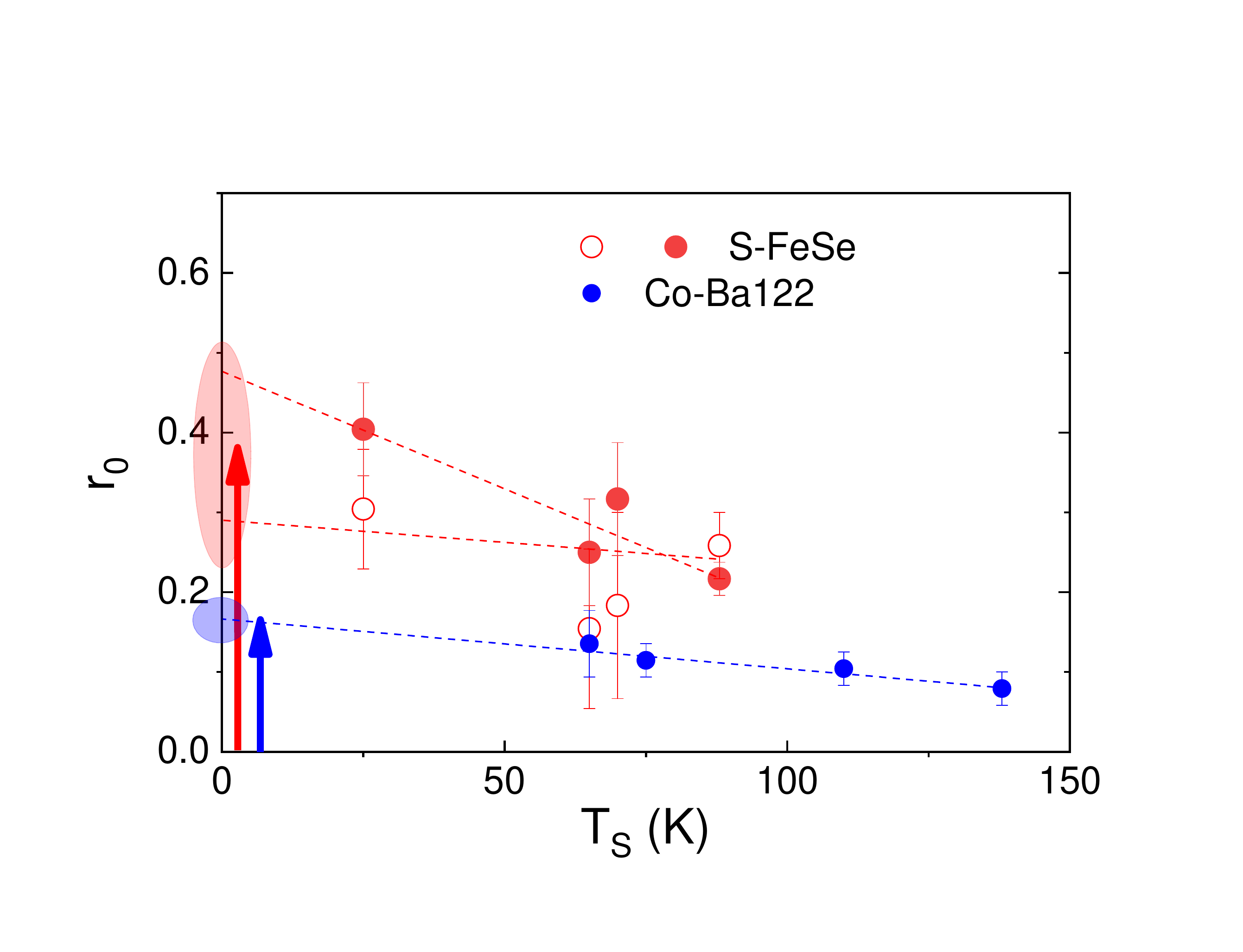}
\caption{\label{fig4} {\bf Lattice-shifted nematic quantum criticality} Dimensionless nemato-electron-lattice coupling $r_0$=$\frac{k_B(T_s-T_0)}{E_F}$ as a function of $T_s$ for FeSe$_{1-x}$S$_x$ and BaFe$_{1-x}$Co$_x$)$_2$As$_2$ \cite{Gallais2016b}. The Fermi energy $E_F$ was taken to be 20 meV and 40 meV for FeSe$_{1-x}$S$_x$ and BaFe$_{1-x}$Co$_x$)$_2$As$_2$ respectively. For FeSe full (empty) circles correspond to $r_0$ values deduced from $T^{QEP}_0$ ($T_0^{400cm^{-1}}$). By definition the origin of the graph corresponds to the location of the QCP in the absence of any coupling to the lattice ($T_S$=$T_0$), the arrows illustrate the effect of the lattice on the QCP. The line are linear fit of $r_0$($T_s$) used to deduced $r_0$($T_S$=0), and the filled ovals illustrate their associated uncertainties.}
\end{figure}

\section*{Discussion}
\subsection*{Effect of nemato-elastic coupling on $T_c$ at the QCP}
As stated above, the $x$ evolution of $T_0$ points to a significant impact of lattice effects
on nematic QCP physics. In order to quantify it we introduce
the ratio $r_0$=$\frac{k_B(T_s-T_0)}{E_F}$ where $E_F$ is the Fermi energy \cite{Paul2017}.  $r_0$=0 signifies the
absence of shift of the QCP. On the other hand, $r_0$ $\sim$ 1 indicates a strong impact of the lattice
on the QCP that can manifest itself by for e.g. a restoration of Fermi liquid behavior above
the QCP \cite{Paul2017,Reiss2020}. For pure FeSe the hole and electron pocket are small, resulting
in small $E_F$ of about 20 meV for each xz/yz derived electron pockets according to ARPES \cite{Reiss2017}.
Adding S only results in the appearance of a small inner-hole pocket above $E_F$, but does not
change appreciably the size of the other pockets, at least below $x$ $\sim$ 0.17. Taking 20 meV as an upper bound for $E_F$
for all FeSe$_{1-x}$S$_x$, we plot $r_0$ versus $T_s$ as shown in fig. 5. Despite some scatter in the
values of $r_0$, the plot suggests a significant value of $r_0$ extrapolated at the thermodynamic
QCP where $T_s$=0: from $r_0$=0.35$\pm$ 0.15. Also shown in fig. 5 are values for Ba122 which are
reduced compared to FeSe$_{1-x}$S$_x$, but still significant ($\sim$ 0.2).
\par
Next we discuss the consequences of our findings for the SC near the nematic QCP of FeSe$_{1-x}$S$_x$. Theoretical
works that ignored nemato-elastic coupling have argued that near a nematic QCP, the critical nematic
fluctuations can lead to a boost in the superconducting $T_c$. On the other hand, once this coupling is
taken into account it leads to a crossover Fermi liquid temperature scale $T_{FL}$=$r_0^{3/2}E_F$
below which the critical nematic fluctuations are suppressed~\cite{Paul2017}. In the case of S-FeSe we estimate
$T_{FL} \sim $ 55 ($\pm$ 25) K. Since $T_{FL} > T_c \sim$ 10 K in this system, it is unlikely
that the critical nematic fluctuations play a role in the pairing in FeSe$_{1-x}$S$_x$. For such a situation
Labat et al. found that significant $T_c$ enhancement at the QCP occurs only if the nemato-elastic
coupling is weak enough such that
$r_0 <(U/V)^2$, where $U$ is an effective interaction in the nematic channel
and $V$ is an effective SC pairing attraction due to e.g. spin-fluctuations \cite{Labat2017}.
For Fe SC we expect $V>U$ since pairing is dominated by spin-fluctuations away from QCP.
While there is currently no realistic theoretical evaluation of $V$ and $U$ for FeSe,
since in practice no $T_c$ boost has been noted in S-FeSe, we conclude that the nemato-elastic
coupling is significantly large with $r_0  > (U/V)^2$.
\par

The same analysis applied to Ba(Fe$_{1-x}$Co$_x$)$_2$As$_2$ gives $T_{FL}\sim$ 40 ($\pm$ 10) K, much closer to the actual $T_c$ observed in these systems. The above estimate suggests that Ba122 systems may be more promising candidates for
a significant effect of the nematic QCP on $T_c$. While the presence of a nearby magnetic QCP in most Ba122 complicates the matter, we note the recent observation of a significant enhancement of $T_c$ near a nematic QCP in Ba$_{1-x}$Sr$_x$Ni$_2$As$_2$ in the absence of any magnetic order \cite{Eckberg2020}. Interestingly, band structure calculations predict a much larger
Fermi surface and therefore a larger $E_F$ with respect to BaFe$_2$As$_2$ \cite{Sudebi2008}. This may possibly lead to a significantly reduced $r_0$, and thus much weaker lattice effects on the QCP of Ba$_{1-x}$Sr$_x$Ni$_2$Fe$_2$. Further investigations, in particular nematic susceptibility measurements, are needed to assess the electron-lattice coupling in this system and confirm this scenario.

\subsection*{Janus face nematicity: itinerant carriers versus local moments}

Finally we address the origin the non-critical HE peak component observed in the $B_{1g}$ spectrum. As discussed above, it only appears in the $B_{1g}$ nematic channel and its energy, around 400 cm$^{-1}$ (50 meV), is only mildly temperature and $x$ dependent. A first possible interpretation is that it comes from interband transitions between the two hole bands centered at the $\Gamma$ point which are split by spin-orbit coupling \cite{Udina2020}. In that case however, we would expect the energy of the HE peak to be very close to the spin-orbit induced energy splitting \cite{Udina2020}, which is estimated from ARPES measurements to be 20 meV \cite{Watson2015}, significantly below 50 meV. In addition, simple calculations of the interband Raman spectrum indicate that it should appear in both $B_{1g}$ and $B_{2g}$ symmetries with comparable intensity, again in disagreement with experiments (see \cite{SM} for Raman calculations of the contribution of interband transitions in the presence of spin-orbit coupling). This leads us to conclude that the HE peak is unlikely of interband origin.
\par
As first argued by Baum et al. \cite{Baum2019} (see also \cite{Glazmada2019}), the HE peak more likely arises from spin fluctuations associated to local moments. Local moments have been argued to be relevant for Fe SC due to Hund's coupling physics \cite{Yin2011,Georges2011,Lanata2013}, where electrons have both itinerant and localized, or "Janus face", character. Numerical calculations of the Raman response within a spin-1 Heisenberg localized spin model with bi-quadratic interactions with exchange parameters believed to be adequate for FeSe have indeed shown the presence of a paramagnon peak in the $B_{1g}$ symmetry channel. Compared to two magnon excitations of ordered antiferromagnets (AF), in FeSe the paramagnon is pushed below typical exchange energies due to magnetic frustration and the presence of both stripe-like and N\'e\'el -like AF fluctuations \cite{Ruiz2020}. The presence of low-lying paramagnon excitations in FeSe has been confirmed by Resonant Inelastic Xray Scattering (RIXS) measurements, where a dispersive collective mode was observed, and interpreted as a paramagnon excitation \cite{Rahn2019}. The energy of the RIXS paramagnon peak near the zone center is close to 50 meV, similar to the Raman HE peak, further enforcing the interpretation of the HE peak in terms of local spin physics. We note that within the Fleury-Loudon formalism of Raman scattering in insulating magnets with nearest neighbors exchange, the $B_{1g}$ Raman scattering operator for spin 1/2 is equivalent to the spin-nematic order parameter written in real space $O_{B_{1g}}$=$P_x-P_y$ where $P_\alpha=\sum_r\textbf{S}_r.\textbf{S}_{r+\alpha}$ \cite{Fleury1968}. For spin 1 systems with bi-quadratic couplings, higher-order terms such as spin quadrupole will also come into play, and may contribute to the $B_{1g}$ HE peak too \cite{Michaud2011,Hu2020}. If interpreted along these lines, the nematic susceptibility extracted from the HE component can be interpreted as the local spin-nematic component of the nematic susceptibility.
\par
In conclusion, the picture emerging from these considerations is that of two distinct component in the nematic fluctuations spectrum in FeSe : one associated with itinerant carrier and displaying clear critical behavior as embodied by $\chi^0_{QEP}$, and the other associated  more correlated electrons and local moment physics which, while sizable, is essentially non-critical as observed for $\chi^0_{HE}$. This suggests that the nematic instability is driven mainly by itinerant carriers. However as we have shown the coupling of these fluctuations to the lattice most likely prevent them from playing a role in enhancing $T_c$ near the nematic QCP of FeSe$_{1-x}$S$_x$. Other Fe SC systems with weaker nemato-elastic coupling, but ideally no other ordering, may provide a better platform to demonstrate this effect. Another open question is whether the local spin component has a role in the superconducting instability in FeSe. Our work suggests that this "Janus face" aspect of nematicity might be a generic feature of Hund's metals like FeSe.

\section*{Methods}
Raman experiments have been carried out using a triple grating JY-T64000 spectrometer. Spectra below 500cm$^{-1}$ were obtained using subtractive mode and 1800 grooves/mm gratings. High-energy spectra between 50 and 2500 cm$^{-1}$ were obtained using a single stage spectrometer with a 600 grooves/mm grating and long-pass edge filter to block the stray light. The spectrometer were equipped with a nitrogen cooled back illuminated CCD detector. All measurement were performed using the 532 nm excitation line from a diode pump solid state laser. Measurement in $B_{1g}$ symmetry channels were obtained using cross incoming and outgoing linear polarization oriented at 45 degrees of the Fe-Fe bonds, while $A_{1g}$+$B_{2g}$ channel were obtained using parallel polarization at 45 degrees of the Fe-Fe bonds. For several $x$ complimentary measurements were performed in the $B_{2g}$ symmetry channel using cross polarization along the Fe-Fe bonds. The laser heating was estimated  in-situ by tracking the onset of elastic light scattering by orthorhombic domains across $T_s$ at different laser powers for $x$=0, $x$=0.05 and $x$=0.14 single crystals, yielding 1K/mW ($\pm$ 0.2) for all samples \cite{Massat2016}. This value was taken for all the other samples. All the raw spectra have been corrected for the Bose factor and the instrumental spectral response. They are thus proportional to the imaginary part of the Raman response function $\chi''(\omega, T)$ in the corresponding symmetry channel.

\subsection*{Samples}
The $x$=0 crystal was grown in Grenoble using the chemical vapor transport method based on the use of an eutectic mixture of AlCl$_3$/KCl as described in \cite{Karlsson2015}.The $x=0.05$, $x=0.07$, $x$=0.14, $x$=0.17 and $x$=0.22 single crystals were grown in Tohoku University by the molten salt flux methods \cite{Urata2016b}. Polycrystalline FeSe$_{1-x}$S$_x$  for the single crystal growth precursors were synthesized by the solid-state reaction methods \cite{Mizoguchi2009}. Polycrystalline FeSe$_{1-x}$S$_x$ and KCl/AlCl$_3$  were mixed up inside the Ar grove box with the molar ratio of 1:12 and were sealed in the quartz tube under the vacuum condition of 10$^{-2}$ Pa. The quartz ampoule was heated using the tube furnace.  The temperature of
hot and cold positions was kept at 390 and 240 $^\circ$C, respectively. After $\sim$ 10 days, single crystals were grown around
the cold part of the quartz tube.
\par
The $x$=0.27 crystal was grown in Ames out of an eutectic mix of KCl/AlCl$_3$ salts as described in \cite{Bohmer2016}.  Details about the growth and characterization of $x$=0.27 single crystals from the same batch can also be found in \cite{Wiecki2018})
\par
 Single crystals from the same batch were characterized by transport measurements yielding first estimates of $T_s$ and $T_c$. The measured crystals were further characterized by SQUID magnetometry to obtain their $T_c$, by EDS and X-ray diffraction to obtain their sulfur content $x$. 
 \par
 X-ray diffraction (XRD) data on FeSe$_{1-x}$S$_{x}$ single crystals were collected at 295 K on a STOE imaging plate diffraction system (IPDS-2T) using Mo $K_{\alpha}$ radiation. All accessible symmetry-equivalent reflections ($\approx 4500$) were measured up to a maximum angle of $2 \Theta =65\deg $. The data were corrected for Lorentz, polarization, extinction, and absorption effects. Using SHELXL \cite{Sheldrick} and JANA2006 \cite{Petricek}, around 101 averaged symmetry-independent reflections ($I > 2 \sigma$) have been included for the refinements in space group (SG) $P4/nmm$.\@ The refinements converged quite well and the somewhat increased reliability factors (see GOF, $R_1$,\@ and $wR_2$ in the Table) and uncertainties for the atomic positions and the ADPs result from the significant mosaic spread often observed for FeSe$_{1-x}$S$_{x}$ samples.\@ Results for $x = 0.12$, $0.22$, and $0.27$ are shown in the supplementary table 1 \cite{SM} as representatives. 
 \par
 In general both EDS and X-ray yielded consistent results within 20$\%$ for $x$ except for the highest S content $x$=0.27. We note that EDS is only a semi-quantitative measure of $x$ unless the element specific X-ray yield are quantitatively calibrated for the given instrument. Therefore XRD data were trusted in most case, except for $x$=0.14 where the value of $T_s$ under the laser spot (25K) indicates a slightly higher S content than the XRD value ($x$=0.12(3)).
 \par
  In addition $T_s$ was estimated in-situ on the same single crystals by monitoring the onset for elastic light scattering by orthorhombic domains at very small powers (0.1 mW) for $x$=0, $x$=0.05 and $x$=0.14. Except for $x$=0.07, $T_s$ value quoted in the text were extracted from this method. For $x$=0.07, transport value from samples from the same batch were taken (see supplementary fig. 1).

\section*{References and Notes}

\section*{Acknowledgments} We thank A. V. Chubukov and T. Shibauchi for fruitful discussions. We also thank G. Wang for his assistance with the EDS measurements. We also thank Laboratoire de Physique des Solides, Universit\'e Paris Saclay for making their MPMS SQUID magnetometer available. Crystal growth and characterization in Ames was supported by the U.S. Department of Energy (DOE), Office of Basic Energy Sciences, Division of Materials Sciences and Engineering. Ames Laboratory is operated for the U.S. DOE by Iowa State University under Contract No. DE-AC02-07CH11358. The contribution from M. M. was supported by the Karlsruhe Nano Micro Facility (KNMF).

\section*{Author Contributions} S. C., P. M. and D. F. performed the Raman scattering experiments with the help of M. C., A. S. and Y. G.  S. C. performed the data analysis and prepared the figures. T. U, Y. T., K. T., S. K., P. S., P. T., A. B. and P. C. grew the single crystals. S. C. performed EDS and SQUID magnetometry measurement and M. M.. performed single crystals XRD to characterize the samples. I. P. provided theoretical insights. Y. G. supervised the project and wrote the paper with inputs from all the authors.

\section*{Competing Interests}
The authors declare no competing interests

\newpage


\end{document}